\documentstyle[prl,preprint,eqsecnum,aps,epsf]{revtex}
\begin{document}
\preprint{KNTP-99-02}
\bibliographystyle{plain}
\title{Winding Number Transition at Finite Temperature : Mottola-Wipf model 
with and without
 Skyrme term}
\author{D. K. Park$^a$, Hungsoo Kim$^b$, Soo-Young Lee$^a$}

\address{$^a$ Department of Physics, Kyungnam University,
Masan, 631-701, Korea.\\
$^b$ Department of Physics, Korea Advanced Institute of Science and
Technology, \\Taejon, 305-701, Korea.}
\date{\today}
 \maketitle

\begin{abstract}
The winding number transition in the Mottola-Wipf model with
and without Skyrme term is examined. For the model with Skyrme term
the number of discrete modes of the fluctuation operator around sphaleron
is shown to be dependent on the value of $\lambda m^2$. Following
Gorokhov and Blatter we derive a sufficient condition for the 
sharp first-order transition, which indicates that first-order transition
occurs when $0< \lambda m^2 < 0.0399$ and $2.148 < \lambda m^2$.
In the intermediate region of $\lambda m^2$ the winding number 
transition is conjectured to be smooth second order. For the model
without Skyrme term the winding number transition is always first order
regardless of the value of parameter. 
\end{abstract}
% \tableofcontents
% \listoffigures
\newpage
% %
\section{Introduction}
Quantum tunneling is one of the most fascinating phenomenon arised due to 
pure quantum
effect and has profound physical implications in many fundamental 
phenomena in various branches of physics. 
After Langer\cite{lan69} and Coleman\cite{col77} opened a
door of the computational methodology for the investigation of the tunneling 
using a classical Euclidean 
solution called instanton or bounce, its generalization 
to the finite temperature tunneling or thermal activation is one of 
the long-standing subject in this field. 
It is well-known that as temperature 
increases from zero temperature 
the pure quantum tunneling at low temperature
is gradually changed to the thermal activation at high 
temperature via thermally assisted tunneling region. This means there is a 
phase transition
between sphaleron\cite{man83}-dominated high temperature regime and the 
instanton-dominated low temperature regime.
This phase transition was explored within quantum mechanical models by 
Affleck\cite{aff81}
and cosmological models by Linde\cite{lin83} about two decades ago. 
Both of them 
argued that 
there exist periodic solutions\cite{kh91} which govern the tunneling 
at intermediate temperature
regime and interpolate smoothly between vacuum instanton and 
sphaleron.
Using a
terminology of the statistical mechanics this phenomenon can be referred to 
second-order transition.
  
Chudnovsky\cite{ch92}, however, showed that this argument is not generic.
Using a simple quantum mechanical model, he derived a relation
\begin{equation}
\frac{d S_E}{d \tau} = E \, > \,0,
\label{sdt}
\end{equation}
where $S_E$ and $\tau$ are Euclidean classical action and period of
periodic solution, respectively. Eq.(\ref{sdt}) allows one to determine
the temperature($T=1/\tau$)-dependence of action($S_E$) from 
energy($E$)-dependence of period($\tau$). In fact, this is completely
analogous to the plot of free enthalphy-vs-pressure of a van der Waals
gas whose equation of state is determined as pressure-vs-volume.
If, for example, $E$-dependence of $\tau$ is monotonic function,
$T$-dependence of $S_E$ becomes also monotonic, and eventually
is merged to the action for the sphaleron solution smoothly. Since
decay rate is proportional to $e^{-S_E}$, this case can be reffered
to second-order transition. If, however, $E$-dependence of $\tau$ is
not monotonic and has one minimum, $T$-dependence of $S_E$ can be
 double-valued function in some region of domain of definition and it
results in a sharp crossover from instanton-dominated low temperature
regime to sphaleron-dominated high temperature regime. This is an 
example of first-order transition. 
Recently, his idea is realized at the 
spin tunneling system\cite{li98} in condensed matter physics. 

Subsequently, Gorokhov and Blatter\cite{go97} extended Chudnovsky's idea
and derived a sufficient condition for the first-order transition using 
only small fluctuation around sphaleron.
Recently, their idea is extended\cite{mu99} for the application to the
model when mass is position-dependent.
 Although they derived the criterion 
to apply it to the some fields of the condensed matter 
physics\cite{go97,go98}, its application to the covariant field theories  is
straightforward. In fact, this criterion is applied to the four-dimensional
scalar field theory with a asymmetric double well potential and the lower 
and upper bounds of the critical value of symmetry breaking parameter is 
calculated\cite{hu99}. Recently, same criterion with that of 
Gorokhov and Blatter is derived again through completely different 
point of view by counting carefully
the number of negative modes for the full hessian of periodic solution near 
sphaleron\cite{soo99}.

In this paper we will examine the winding number transition of the 
Mottola-Wipf(MW) model\cite{mo89} with and without Skyrme term\cite{sk61}
and clarify the type of the transition in the full range of parameter
space. Since the method of Ref.\cite{soo99} for the study of winding 
number transition needs a periodic instanton solution which is 
very complicated\cite{hab96} in this model, we will follow the procedure of 
Ref.\cite{go97}. 

MW model, non-linear $O(3)$ model with a soft symmetry breaking term, is usually adopted
as a toy model for the study of the baryon number violation in electroweak
theory\cite{ya89}. Although localized vacuum instanton does not exist
in this model, the analytical expression of the sphaleron solution
was derived by paralleling Manton's original argument in Ref.\cite{mo89}.
The reason why MW model can be used as a toy model for a investigation of 
baryon-number violation in the electroweak theory is nicely explained in 
Ref.\cite{mo89}. For a completeness we summarize it briefly. If one defines
two component field
\begin{equation}
 \psi = \left( \begin{array}{c}
                \psi_1 \\
                \psi_2 
                \end{array}
          \right) 
\end{equation}
such that $\phi^a = \psi^{\dagger} \sigma^a \psi$ where $phi^a$ is scalar field of
MW model and  $\sigma^a$ is usual 
Pauli matrix. Introducing gauge field $A_{\mu} = (\psi^{\dagger} \partial_{\mu}
\psi - \partial_{\mu} \psi^{\dagger} \psi)/2i$, one can show that the 
winding number $Q$ of MW model which is defined as
\begin{equation}
Q = \frac{1}{8\pi} \int d^2 x  \epsilon_{\mu \nu} \vec{\phi} \cdot 
(\partial_{\mu} \vec{\phi} \times \partial_{\nu} \vec{\phi} )
\end{equation}
is reduced to 
\begin{equation}
\frac{1}{4\pi} \int d^2 x \epsilon_{\mu \nu} F_{\mu \nu}.
\label{fmn}
\end{equation}
The integrand in Eq.(\ref{fmn}) is nothing but an chiral anomaly\cite{ja85} of 
massless fermion field coupled to $U(1)$ gauge field.
Through index theorem the transition in MW model can be interpreted as 
tunneling which connects states with different chiral fermion numbers.

In order to make a modified model which supports a localized vacuum 
instanton as 
well as sphaleron, Piette {\it et al}\cite{pi94} added a
Skyrme term to this model. 
In this modified model whose action and constraint are
\begin{eqnarray}
\label{action1}
S&=& \frac{1}{g^2} \int d\tau dx 
\left[ \frac{1}{2} \partial_{\mu} \phi^{a} \partial_{\mu} \phi^a
       + m^2 (1 + \phi^3) + \frac{\lambda}{8}
       \phi_{\mu \nu}^{ab} \phi_{\mu \nu}^{ab} \right]  \\  \nonumber  
& & \hspace{1.0cm} \phi^a \phi^a = 1,\hspace{.5cm}   a = 1, 2, 3 \hspace{.5cm}  \mu, \nu = \tau, x
%\label{action1}
\end{eqnarray}
where $\phi_{\mu \nu}^{ab} = \partial_{\mu} \phi^a \partial_{\nu} \phi^b
- \partial_{\nu} \phi^a \partial_{\mu} \phi^b$, sphaleron is identical to
that of MW model whose action is Eq.(\ref{action1}) with $\lambda = 0$ and 
localized instanton can be obtained numerically. Winding number transition
between instanton-dominated and sphaleron-dominated regimes in this model
are discussed at particulur value of parameter, $\lambda m^2 = 0.001$,
in Ref.\cite{ku97}. In this paper we will discuss the winding number
transition in the full
range of parameter space. 

The paper is organized as follows. In Sec.II we will briefly review 
Gorokhov and Blatter's procedure in covariant scalar field theory. In 
Sec.III we will expand the field equation of 
Mottola-Wipf-Skyrme(MWS) model around sphaleron solution by introducing
the fluctuation fields $u$ and $v$. The spectra of the spatial fluctuation
operators $\hat{h}_u$ and $\hat{h}_v$ are discussed at Sec.IV. In Sec.V,
following Gorokhov and Blatter, we will examine the winding number transition
in MWS model. It will be shown that the type of transition is either
first-order or second-order depending on the value of $\lambda m^2$.
In Sec.VI the winding number transition in MW model will be discussed.
It will be shown that the type of transition in this model is always 
first order regardless of value of parameter. In Sec.VII a brief conclusion
will be given.

\section{Criterion for sharp first-order transition} 
In this section we will review the procedure of Ref.\cite{go97} briefly
in the covariant field theory. The Euclidean action we will consider is 
\begin{equation}
S = \int dx 
\left[ \frac{1}{2} \partial_{\mu} \phi \partial_{\mu} \phi + V(\phi) \right]
\end{equation}
where field equation is 
\begin{equation}
\Box \phi = V^{\prime}(\phi).
\end{equation}
Although we do not need a specific form of $V(\phi)$, we assume field equation
allows periodic instanton and sphaleron configurations. Since sphaleron
$\phi_{sph}$ is static configuration, its equation of motion reads
\begin{equation}
\frac{\partial^2 \phi_{sph}}{\partial x^2} = V^{\prime} (\phi_{sph}).
\end{equation}

Now, we consider a small fluctuation $\eta(x, \tau)$ around sphaleron
solution. If one uses Taylor expansion in potential, it is easy to show
$\eta(x, \tau)$ satisfies upto the order of $\eta^3$
\begin{equation}
\label{expand1}
\hat{l} \eta(x, \tau) = \hat{h} \eta(x, \tau) + G_2[\eta] + G_3[\eta]
%\label{expand1}
\end{equation}
where $\hat{l} = \frac{\partial^2}{\partial \tau^2}$, 
$\hat{h} = - \frac{\partial^2}{\partial x^2} + V^{\prime \prime} (\phi_{sph})$,
$G_2[\eta] = \frac{1}{2} V^{\prime \prime \prime} (\phi_{sph}) \eta^2$, 
and $G_3[\eta] = \frac{1}{6} V^{\prime \prime \prime \prime} (\phi_{sph})
\eta^3$. The main idea of Ref.\cite{go97} is  to compare the frequency of $\eta(x, \tau)$ 
with that of sphaleron by solving Eq.(\ref{expand1})  perturbatively,
 which yields a sufficient condition for the first-order
transition.
Here, the frequency of sphaleron $\omega_{sph}$ means frequency of the periodic
instanton solution when periodic instanton approaches to the sphaleron solution.

To the lowest order in perturbation we use {\it ansatz}
\begin{equation}
\label{lowest}
\eta(x, \tau) = a u_0(x) \cos \omega_{sph} \tau.
%\label{lowest}
\end{equation}
Here, $a$ is a small parameter, which is associated with small amplitude of 
periodic solution whose center is $\phi_{sph}$ at quantum mechanical model.
Substituting Eq.(\ref{lowest}) into Eq.(\ref{expand1}) while 
neglecting terms of order
higher than $a$ one obtains
\begin{equation}
\hat{h} u_0(x) = l(\omega_{sph}) u_0(x)
\end{equation}
where $l(\omega) \equiv  - \omega^2$. Since spatial operator $\hat{h}$ has only
one negative mode and $l(\omega_{sph}) < 0$, $u_0(x)$ and $l(\omega_{sph})$
are nothing but the ground state eigenfunction and eigenvalue of $\hat{h}$,
respectively. This fact allows one to calculate $\omega_{sph}$ by solving
the spectrum of $\hat{h}$. 

To next order perturbation we use
\begin{equation}
\label{first}
\eta(x, \tau) = a u_0(x) \cos \omega \tau + a^2 \eta_1(x, \tau)
%\label{first}
\end{equation}
where $\omega$ is the correction to the frequency $\omega_{sph}$. Inserting
Eq.(\ref{first}) into Eq.(\ref{expand1}) it is easy to obtain
\begin{equation}
\label{first1}
\eta_1 = (\hat{l} - \hat{h})^{-1} \chi_1
%\label{first1}
\end{equation}
where
\begin{equation}
\chi_1 = \frac{1}{a}
\left[ l(\omega_{sph}) - l(\omega) \right] u_0(x) \cos \omega \tau
+ G_2[u_0] \cos^2 \omega \tau.
\end{equation}
Since $\hat{l} - \hat{h}$ has zero mode $\mid u_0(x) \cos \omega \tau >$,
we can escape infinity in Eq.(\ref{first1}) if and only if a condition 
$<u_0(x) \cos \omega \tau \mid \chi_1 > = 0$ holds. This condition requires
no shift in the frequency $\omega = \omega_{sph}$, which means that 
next order perturbation must be performed to find frequency shift. 
Before we go to the next order, it is in order to compute $\eta_1$ explicitly.
It is straightforwardly achieved from Eq.(\ref{first1}):
\begin{equation}
\eta_1(x, \tau) = g_1(x) + g_2(x) \cos 2 \omega_{sph} \tau
\end{equation}
where
\begin{eqnarray}
g_1(x)&=&- \frac{1}{2} \hat{h}^{-1} G_2[u_0]   \\  \nonumber
g_2(x)&=&- \frac{1}{2} [\hat{h} - l(2 \omega_{sph})]^{-1} G_2[u_0].
\end{eqnarray}

Next order perturbation can be done by using
\begin{equation}
\label{second}
\eta(x, \tau) = a u_0(x) \cos \omega \tau + a^2 \eta_1(x, \tau)
                + a^3 \eta_2(x, \tau).
%\label{second}
\end{equation}
Substituting Eq.(\ref{second}) into Eq.(\ref{expand1}) and 
neglecting terms of order $a^4$
and higher, one gets
\begin{equation}
\eta_2(x, \tau) = a^{-3} (\hat{l} - \hat{h})^{-1}
\left[ \chi^{(0)}_2 + \chi^{(1)}_2 \cos \omega \tau + 
       \chi^{(2)}_2 \cos 2 \omega \tau + \chi^{(3)}_2 \cos 3 \omega \tau
                                                              \right]
\end{equation}
where
\begin{eqnarray}
\chi^{(0)}_2&=&a^2 \left[ \hat{h} g_1(x) + \frac{1}{2} G_2[u_0] \right]  \\ 
                                                                    \nonumber
\chi^{(1)}_2&=&-a \left[ l(\omega) - l(\omega_{sph}) \right] u_0(x) \\ 
                                                                   \nonumber
            & & +a^3 \left[ \frac{\partial G_2[\xi]}{\partial \xi} 
	    \bigg |_{\xi = u_0} \left(g_1(x) + \frac{1}{2} g_2(x)\right)
	    + \frac{3}{4} G_3[u_0]   \right]  \\ \nonumber 
\chi^{(2)}_2&=&a^2 \left[ [\hat{h} - l(2 \omega)] g_2(x) + \frac{1}{2}
                          G_2[u_0]   \right]    \\   \nonumber
\chi^{(3)}_2&=&a^3 \left[ \frac{1}{2} \frac{\partial G_2[\xi]}{\partial \xi}
            \bigg |_{\xi = u_0} g_2(x) + \frac{1}{4} G_3[u_0] \right].
\end{eqnarray}

The infinity problem mentioned above yields another condition
$<u_0(x) \mid \chi^{(1)}_2> = 0$, which results in
\begin{equation}
l(\omega) - l(\omega_{sph}) = 
- \frac{a^2}{2 <u_0 \mid u_0>} <u_0 \mid f[u_0] >
\end{equation}
where
\begin{equation}
f[u_0] = \frac{\partial G_2[\xi]}{\partial \xi} \bigg |_{\xi = u_0}
\left[ \hat{h}^{-1} + \frac{1}{2} [\hat{h} - l(2 \omega_{sph})]^{-1} \right]
G_2[u_0] - \frac{3}{2} G_3[u_0].
\end{equation}
If the spectrum of $\hat{h}$, $\hat{h} \mid u_n> = h_n \mid u_n>$, is 
completely solved, $<u_0 \mid f[u_0]>$ is reduced to 
%\begin{eqnarray}
\begin{equation}
<u_0 \mid f[u_0]> 
= 2 \sum_n \left( \frac{1}{h_n} + \frac{1}{2} \frac{1}{h_n -
                                      l(2 \omega_{sph})}  \right)
\mid <G_2[u_0] \mid u_n> \mid^2
- \frac{3}{2} <u_0 \mid G_3[u_0]>.
\label{uf}
\end{equation}
%\end{eqnarray}
One may worry for the use of Eq.(\ref{uf}) due to zero mode of $\hat{h}$, 
say $h_1=0$. It is, however, easy to show that it does not cause any problem
if one realizes $< G_2 [ u_0] | u_1 > = 0 $ because of odd parity of $u_1$.

Before we derive a sufficient condition for the sharp transition, it 
may be helpful to comment on the two kinds of field theoretical 
tunneling models. 
One kind(case I) consists of models which support localized vacuum
instanton and sphaleron. The usual quantum mechanical model and MWS
model plunge into this case. 
In these models the first-order phase transition takes place as usual
quantum mechanical cases. The typical figure of first-order transition
in case I is given in Fig.1(a).
Another kind(case II) consists of models which
does not support localized vacuum instanton. The MW model and electroweak
theory are categorized in this case. 
In these models the low energy periodic instanton is built out of constraint
instanton\cite{kh91} whose period goes to zero at $E \rightarrow 0$. Applying
a relation (\ref{sdt}) one can get a typical figure of first-order transition
in case II which is shown in Fig.1(b).
In both cases the first-order transition occurs when $\omega > \omega_{sph}$,
which results in 
\begin{equation}
<u_0 \mid f[u_0]> \hspace{.3cm} > \hspace{.3cm} 0.
\end{equation}
We will use this criterion to determine the types of transition in the 
full range of parameter space in MWS and MW models.

\section{Fluctuation around sphaleron in MWS model}
The field equation of MWS model whose action is given at Eq.(\ref{action1}) 
is easily
derived by introducing and subsequently removing a Lagrange multiplier:
\begin{equation}
(\delta^{ab} - \phi^a \phi^b)
\left[ \Box \phi^b + \lambda \partial_{\mu} (\phi^{bc}_{\mu \nu} 
      \partial_{\nu} \phi^c)  \right]
= m^2 (\delta^{a3} - \phi^a \phi^3).
\end{equation}
Since Skyrme term does not contribute to field equation for static
sphaleron\cite{pi94}, sphaleron is same with that of MW model which is 
derived at Ref.\cite{mo89}:
\begin{equation}
\phi_{sph} = (\sin \xi_{sph}, 0, - \cos \xi_{sph})
\end{equation}
where 
\begin{equation}
\xi_{sph} = 2 \sin^{-1} ({\rm sech} m x).
\end{equation}

Now, we introduce small fluctuation fields $u$ and $v$ around sphaleron
such as
\begin{equation}
\phi = \frac{1}{\sqrt{1 + u^2}}
\left(   \sin (\xi_{sph} + v), u, -\cos (\xi_{sph} + v)  \right).
\end{equation}
The expansion of the field equation (3.1) in terms of $u$ and $v$ is very
tedious, but straightforward:
\begin{equation}
\label{expand2}
\hat{l} \left( \begin{array}{c}
               u \\ v
	       \end{array}  \right)
= \hat{h} \left( \begin{array}{c}
               u \\ v
	         \end{array}  \right)
+ \left( \begin{array}{c}
          G^u_2(u, v) \\ G^v_2(u, v)
	  \end{array}   \right)
+ \left( \begin{array}{c}
          G^u_3(u, v) \\ G^v_3(u, v)
	 \end{array}   \right)
+ \cdots
%\label{expand2}
\end{equation}
where
\begin{eqnarray}
\hat{l}&=&\left( \begin{array}{clcr}
                 \frac{\partial^2}{\partial \tau^2} & 0   \\
		 0 & \frac{\partial^2}{\partial \tau^2}  
		 \end{array}     \right)    
                                            %  \\   \nonumber
\hspace{2.0cm}
\hat{h}=\left( \begin{array}{clcr}
                 \hat{h}_u & 0    \\
		 0 & \hat{h}_v
		 \end{array}     \right)   \\   \nonumber
G^u_2(u, v)&=& \frac{{\rm sech} m x}{1 + 4 \lambda m^2 {\rm sech}^2 
                                                                      m x}
              \left[2 m^2 \tanh m x (u v - \lambda \dot{u} \dot{v})
	            - 4 m u v^{\prime} + 2 \lambda m 
		    (2 \dot{u}^{\prime} \dot{v} - 2 \ddot{u} v^{\prime}
		     + u^{\prime} \ddot{v} - \dot{u} \dot{v}^{\prime})
	                                                          \right]
								  \\ \nonumber
G^v_2(u, v)&=& {\rm sech} m x
              \left[ - m^2 \tanh m x (u^2 - v^2 - 2 \lambda 
	            \dot{u}^2) + 4 m u u^{\prime}
		    + 2 \lambda m (\ddot{u} u^{\prime} - \dot{u} 
		                         \dot{u}^{\prime} )
					                    \right]  \\ 
                                                                    \nonumber
G^u_3(u, v)&=& \frac{1}{1 + 4 \lambda m^2 {\rm sech}^2 m x}
              \Bigg[ 2 u (\dot{u}^2 + u^{\prime 2}) 
	             - u (\dot{v}^2 + v^{\prime 2})
		     -\frac{m^2}{2} ( 1 - 2 {\rm sech}^2 m x)
		     (u v^2 - u^3)    \\  \nonumber    
	 & & \hspace{3.5cm}            + \lambda \bigg[
		                     -\ddot{u} v^{\prime 2} 
				     -\dot{u} v^{\prime} \dot{v}^{\prime}
				     +2 \dot{u}^{\prime} \dot{v} v^{\prime}
				     + u^{\prime} \ddot{v} v^{\prime}
				     -u^{\prime} \dot{v} \dot{v}^{\prime} 
				                       \\  \nonumber
        & & \hspace{4.5cm}	     +\dot{u} \dot{v} v^{\prime \prime}
				     -u^{\prime \prime} \dot{v}^2
				     + 4 m^2 {\rm sech}^2 m x
				     (3 u \dot{u}^2 + u^2 \ddot{u}) \bigg]
				                                     \Bigg]
								\\ \nonumber
G^v_3(u, v)&=& 2 u(\dot{u} \dot{v} + u^{\prime} v^{\prime})
               + \frac{m^2}{6} (1 - 2 {\rm sech}^2 m x)
	         (3 u^2 v - v^3)    \\   \nonumber
           & & \hspace{1.0cm} + \lambda 
                             \left[ - \dot{u} \dot{u}^{\prime} v^{\prime}
	                      -\dot{u}^2 v^{\prime \prime}
			      - u^{\prime 2} \ddot{v} 
			      -\dot{u}^{\prime} u^{\prime} \dot{v}
			      +2 \dot{u} u^{\prime} \dot{v}^{\prime}
			      + \dot{u} u^{\prime \prime} \dot{v}
			      + \ddot{u} u^{\prime} v^{\prime}
			                                       \right].
							     \nonumber
\end{eqnarray}
Here, prime and dot mean differentiation with respect to $x$ and $\tau$
respectively, and $\hat{h}_u$ and $\hat{h}_v$ are
\begin{eqnarray}
\hat{h}_u&=&\frac{1}{1 + 4 \lambda m^2 {\rm sech}^2 m x}
         \left[ - \frac{\partial^2}{\partial x^2}
	        + m^2(1 - 6 {\rm sech}^2 m x) \right]
		                                      \\  \nonumber
\hat{h}_v&=& - \frac{\partial^2}{\partial x^2} + 
              m^2 (1 - 2 {\rm sech}^2 m x).
\end{eqnarray}

It is interesting that Skyrme term contributes only modification of $\hat{h}_u$,
not $\hat{h}_v$. Apart from denominator in $\hat{h}_u$ the eigenvalue equations
of $\hat{h}_u$ and $\hat{h}_v$ are usual P\"{o}schl-Teller type which can be
solved completely\cite{ja77}. The denominator in $\hat{h}_u$ modifies the 
spectrum which will be discussed in next section.

\section{Spectra of \lowercase{$\hat{h}_u$} and \lowercase{$\hat{h}_v$}}
Consider an eigenvalue equation of the usual P\"{o}schl-Teller type 
potential:
\begin{equation}
\label{teller}
\left[ - \frac{d^2}{d x^2} + V(x) \right] \psi_n(x) = \lambda_n \psi_n(x)
%\label{teller}
\end{equation}
where
\begin{equation}
V(x) = \omega^2 - \frac{V_0}{\cosh^2 \omega x}.
\end{equation}
It is well-known that Eq.(\ref{teller}) has both finite discrete modes, whose
number is dependent on 
\begin{equation}
\label{svalue}
s = \frac{1}{2} \left[-1 + \sqrt{1 + \frac{4 V_0}{\omega^2}} \right],
%\label{svalue}
\end{equation}
and continuum states. The spectrum and eigenfunctions are summarized as 
follows\cite{ja77};

(1) Discrete modes
\begin{eqnarray}
\label{discrete}
\lambda_n&=& \omega^2 [1 - (s - n)^2] \hspace{3.0cm} (n = 0, 1, \cdots, 
                                                        n_{max} < s)
							   \\ \nonumber
\psi_n(x)&=& \sqrt{\frac{\omega}{n!}}
            \sqrt{\frac{\Gamma(1+2s-n)}{\Gamma(1+s-n) \Gamma(s-n)}}
	    2^{n-s} \cosh^{n-s} \omega x  \\ \nonumber
         &\times& _2F_1(-n, 1+2s-n; s-n+1; \frac{1}{2} (1 - \tanh \omega x))
%\label{discrete}
\end{eqnarray}

(2) Continuum modes
\begin{eqnarray}
\label{continuum}
\lambda_k&=& \omega^2 + k^2   \\ \nonumber
\psi_k(x)&=& \frac{1}{\sqrt{2 \pi}}
             \frac{\Gamma(-s - \frac{i k}{\omega}) \Gamma(1+s - \frac{i k}
	                                                        {\omega})}
		  {\Gamma(-\frac{i k}{\omega}) \Gamma(1 - \frac{i k}
		                                               {\omega})}
		e^{i k x}      \\ \nonumber
       &\times&	_2F_1(s+1, -s; 1 - \frac{i k}{\omega}; \frac{1}{2}
		                                      (1 - \tanh \omega x)).
%\label{continuum}
\end{eqnarray}
Here, $\Gamma(a)$ and $_2F_1(a, b; c; z)$ are usual gamma and hypergeometric
functions.
Since eigenvalue equation for $\hat{h}_v$, $\hat{h}_v \varphi_n(x) = 
\epsilon^v_n \varphi_n(x)$, is exactly P\"{o}schl-Teller type equation, it is 
very easy to find its spectrum and eigenfunctions, which is 
summarized at Table I.

Now, let us consider eigenvalue equation of $\hat{h}_u$;
\begin{equation}
\label{eigenu}
\hat{h}_u \psi_n(x) = \epsilon^u_n \psi_n(x).
%\label{eigenu}
\end{equation}
After rearranging Eq.(\ref{eigenu}) one can show that it is possible to make
it  
as usual P\"{o}schl-Teller type provided the following explicit
$\epsilon^u_n$-dependence of $s$ is 
allowed ;
\begin{equation}
\label{svalue1}
s = \frac{1}{2} \left[ \sqrt{25 + 16 \lambda \epsilon^u_n} - 1 \right].
%\label{svalue1}
\end{equation}
After eliminating $\epsilon^u_n$ by using Eq.(\ref{discrete}) and 
(\ref{svalue1}) one can arrive 
at conclusion that different $s$-value is assigned to each discrete mode,
\begin{equation}
\label{svalue2}
s_n = \frac{(8 n \lambda m^2 - 1) + 
            \sqrt{64 \lambda^2 m^4 - 16(n^2+n-7) \lambda m^2 + 25}}
	   {2(1 + 4 \lambda m^2)}.
%\label{svalue2}
\end{equation}

Now, we can solve Eq.(\ref{svalue2}) 
for non-negative integer $n$ under requirements 
$s > 0$ and $s > n$, which results in  a fact that the number 
of discrete modes
is dependent on $\lambda m^2$. For example, there are three discrete 
modes($n = 0, 1, 2$) for $0 < \lambda m^2 \leq 3/2$. And, one more
discrete mode appears for $3/2 < \lambda m^2 \leq 7/2$.
Generally, there are $n$ discrete modes for 
$\frac{(n-1)^2 + (n-1)-6}{4} < \lambda m^2 \leq
\frac{n^2+n-6}{4}$. Parameter-dependence of discrete spectrum is explicitly
shown at Fig.2.
It is worthwhile noting that $n=0$ is negative mode and $n=1$ is zero mode
which indicates that $u$ is fluctuation field around unstable configuration.

The continuum spectrum of $\hat{h}_u$ is more simple. If one follows the 
same procedure, it is easy to show that continuum spectrum and correspondent
eigenfunctions are exactly same with those of Eq.(\ref{continuum}) provided
$s$ is replaced by 
\begin{equation}
s_k = \frac{-1 + \sqrt{25 + 16 \lambda (m^2 + k^2)}}{2}.
\end{equation}
Table II shows a complete spectrum of $\hat{h}_u$ for $0 < \lambda m^2
< 3/2$.

\section{Criterion for first-order transition in MWS model}
In this section we will derive the criterion for the first-order transition
in MWS model by following Gorokhov and Blatter. To do the lowest order
perturbation we choose a {\it ansatz}
\begin{equation}
\label{mwslowest}
\left( \begin{array}{c}
        u  \\  v
       \end{array}    \right) = a 
\left( \begin{array}{c}
        u_0(x)  \\  v_0(x)
	\end{array}   \right)  \cos \Omega_{sph} \tau
%\label{mwslowest}
\end{equation}
where $a$ is discussed at the below of Eq.(\ref{lowest}). 
Inserting Eq.(\ref{mwslowest}) into 
Eq.(\ref{expand2}) and neglecting terms of order higher than $a$ we obtain
\begin{equation}
\hat{h} \left( \begin{array}{c}
                u_0(x)  \\ v_0(x)
		\end{array}    \right) = l(\Omega_{sph})
         \left( \begin{array}{c}
	                 u_0(x)  \\ v_0(x)
	         \end{array}    \right)
\end{equation}
where $l(\Omega) \equiv - \Omega^2$. Based on the analysis of Sec. IV, it is easy to
show that $\hat{h}$ has only one negative mode whose eigenvalue and 
eigenfunction are $\epsilon^u_0 = - m^2 (s_0^2 - 1)$ and 
$ \left( \begin{array}{c}
          \psi_0(x)  \\  0
	  \end{array}   \right) $, respectively. Here, $s_0$ is obtained from
Eq.(\ref{svalue2}) by letting $n=0$;
\begin{equation}
s_0 = \frac{\sqrt{1 + 4 (1 + 4 \lambda m^2)(6 + 4 \lambda m^2)} -1}
           {2(1 + 4 \lambda m^2)}.
\end{equation}
Hence, the lowest perturbation shows
\begin{eqnarray}
\Omega_{sph}&=& \sqrt{s_0^2 - 1} m  \\  \nonumber 
u_0(x)&=& \psi_0(x) = 2^{-s_0} \sqrt{\frac{m \Gamma(1 + 2 s_0)}
                                    {\Gamma(s_0) \Gamma(1 + s_0)}}
	                        	\cosh^{-s_0} m x   \\  
v_0(x)&=& 0.  \nonumber
\end{eqnarray}

Now, let us start the next order perturbation by substituting
\begin{equation}
\left( \begin{array}{c}
       u(x, \tau)  \\  v(x, \tau)
       \end{array}                 \right) = 
\left( \begin{array}{c}
       a u_0(x) \cos \Omega \tau + a^2 u_1(x, \tau)  \\
       a^2 v_1(x, \tau)
       \end{array}       \right)
\end{equation}
into Eq.(\ref{expand2}). As commented in Sec.II, $\Omega$ is the correction to the 
frequency $\Omega_{sph}$. Neglecting higher order terms one arrives at the 
following equation:
\begin{equation}
\left( \begin{array}{c}
       u_1(x, \tau)  \\  v_1(x, \tau)
       \end{array}     \right) = 
(\hat{l} - \hat{h})^{-1} \vec{\chi_1}
\end{equation}
where
\begin{equation}
\vec{\chi_1} = \frac{1}{a} [l(\Omega_{sph}) - l(\Omega)]
               \left( \begin{array}{c}
	              u_0(x) \cos \Omega \tau  \\  0
		      \end{array}       \right)
	       + \left( \begin{array}{c}
	                0  \\  H_1(x, \tau)
			\end{array}               \right).
\end{equation}
Here, $H_1(x, \tau)$ is 
\begin{eqnarray}
H_1(x, \tau)&=& 2^{-2 s_0}
              \frac{m^3 \Gamma(1 + 2 s_0)}
	           {\Gamma(s_0) \Gamma(1 + s_0)}
	      \frac{\sinh m x}
	           {\cosh^{2 s_0 + 2} m x}   \\  
& & \times \left[ - (4 s_0 + 1) \cos^2 \Omega \tau + 2 \lambda \Omega^2
                  (\sin^2 \Omega \tau + s_0)  \right].  \nonumber
\end{eqnarray}
Using same argument given at Sec.II, we need a condition
\begin{equation}
\label{cond1}
< \left( \begin{array}{c}
         u_0 \cos \Omega \tau \\ 0
	 \end{array}     \right)     \bigg | \vec{\chi_1} > = 0
%\label{cond1}
\end{equation}
to escape an infinity arised due to inversion of $\hat{l} - \hat{h}$.
The condition (\ref{cond1}) is 
automatically satisfied if $\Omega = \Omega_{sph}$,
which means there is no frequency shift in the present order of  perturbation.

Before we go to next order, let us compute $u_1(x, \tau)$ and $v_1(x, \tau)$
explicitly. This is achieved by direct calculation, which
needs a tedious, but straightforward procedure:
\begin{eqnarray}
u_1(x, \tau)&=& 0   \\ \nonumber
v_1(x, \tau)&=& g_{v, 1}(x) + g_{v, 2}(x) \cos 2 \Omega_{sph} \tau
\end{eqnarray}
where
\begin{eqnarray}
\label{mwsguv}
g_{v, 1}(x)&=&2^{-2s_0} \frac{m^3 \Gamma(1 + 2 s_0)}
                             {\Gamma(s_0) \Gamma(1 + s_0)}
	       \left[2 s_0(1 - \lambda \Omega_{sph}^2) + 
	             (\frac{1}{2} - \lambda \Omega_{sph}^2) \right] \\ \nonumber
           & &\times \hat{h}_v^{-1} \frac{\sinh m x}
	                              {\cosh^{2 s_0 + 2} m x}  \\
                                                                    \nonumber
g_{v, 2}(x)&=&-2^{-2s_0} \frac{m^3 \Gamma(1 + 2 s_0)}
                             {\Gamma(s_0) \Gamma(1 + s_0)}
			     (2 s_0 + \frac{1}{2} + \lambda \Omega_{sph}^2)
			                                          \\ \nonumber
           & &\times \left[l(2 \Omega_{sph}) - \hat{h}_v \right]^{-1}
	   \frac{\sinh m x}{\cosh^{2 s_0 + 2} m x}.
%\label{mwsguv}
\end{eqnarray}

Now, let us perform next order perturbation with 
\begin{equation}
\label{mwsfirst}
\left( \begin{array}{c}
        u(x, \tau)  \\   v(x, \tau)
	\end{array}        \right) = 
\left( \begin{array}{c}
       a u_0(x) \cos \Omega \tau + a^3 u_2(x, \tau)  \\
       a^2 v_1(x, \tau) + a^3 v_2(x, \tau)
       \end{array}          \right).
%\label{mwsfirst}
\end{equation}
After inserting Eq.(\ref{mwsfirst}) into Eq.(\ref{expand2}) we obtain
\begin{equation}
\label{mwsfirst1}
\left( \begin{array}{c}
       u_2(x, \tau)  \\  v_2(x, \tau)
       \end{array}       \right) = (\hat{l} - \hat{h})^{-1} 
       \vec{\chi}_2.
%\label{mwsfirst1}
\end{equation}

Here, we do not need full explicit form of $\vec{\chi}_2$ for the 
derivation of criterion for sharp winding number transition. The only
one we need is the upper element in Eq.(\ref{mwsfirst1}), which is
\begin{equation}
\label{upper1}
u_2(x, \tau) = (\hat{l} - \hat{h}_u)^{-1} \chi_2^u
%\label{upper1}
\end{equation}
where
\begin{eqnarray}
\label{upper2}
\chi_2^u&=& - \frac{1}{a^2}
           \left[ l(\Omega) - l(\Omega_{sph}) \right] u_0(x) \cos \Omega \tau
	                                                             \\
                                                                     \nonumber
        & & + \frac{1}{1 + 4 \lambda m^2 {\rm sech}^2 m x}
	   \left[ G_{u, 1}(x) \cos \Omega \tau + G_{u, 2}(x) \cos 3 
	          \Omega \tau                           \right].
%\label{upper2}
\end{eqnarray}
In Eq.(\ref{upper2}) the explicit form of $G_{u, 2}(x)$ is not needed also
for the derivation of criterion. The only one we need is $G_{u, 1}(x)$
which is 
\begin{eqnarray}
\label{gu1}
G_{u, 1}(x)&=& 2^{-s_0} \left( 
                               \frac{m \Gamma(1 + 2 s_0)}
                                    {\Gamma(s_0) \Gamma(1 + s_0)}
				                   \right)^{\frac{1}{2}}
	   \Bigg[ \frac{2 m^2 \sinh m x}
	               {\cosh^{s_0+2} m x}
		   \left(g_{v, 1} + \frac{1}{2} g_{v, 2} - \lambda
		         \Omega_{sph}^2 g_{v, 2}   \right) 
			                             \\   \nonumber 
                   & &     \hspace{5.0cm}			 - 
	           \frac{4 m}{\cosh^{s_0+1} m x}
		   \left(g_{v, 1}^{\prime} + \frac{1}{2} g_{v, 2}^{\prime}
		         - \lambda \Omega_{sph}^2 g_{v, 1}^{\prime} \right)
			                                       \Bigg]
							   \\ \nonumber
           & &+ 2^{-3 s_0} 
	      \left( \frac{m \Gamma(1 + 2 s_0)}
	                  {\Gamma(s_0) \Gamma(1 + s_0)}
			                    \right)^{\frac{3}{2}}
	  \Bigg[ \left[ \frac{1}{2} \Omega_{sph}^2 + \frac{3}{8} m^2
	                                          (1 + 4 s_0^2)    \right]
	           \cosh^{-3 s_0} m x
		                                \\  \nonumber
& & \hspace{5.0cm}  - \frac{3}{4} m^2(1 + 2 s_0^2) \cosh^{-3s_0-2} m x
		                                             \Bigg].
%\label{gu1}
\end{eqnarray}
Hence, the condition $<u_0(x) \cos \Omega \tau \mid \chi^u_2> = 0$ gives the 
correction to the frequency:
\begin{equation}
\label{fshift1}
l(\Omega) - l(\Omega_{sph}) = a^2
< u_0(x) \mid \frac{\cosh^2 m x}
                   {\cosh^2 m x + 4 \lambda m^2}
	      G_{u, 1}(x) >.
%\label{fshift1}
\end{equation}

In order to change Eq.(\ref{fshift1}) to a more convenient form, we have to 
compute $g_{v, 1}(x)$ and $g_{v, 2}(x)$ explicitly which is performed in 
Appendix, where the following integral representations are obtained:
\begin{eqnarray}
\label{guv12}
g_{v, 1}(x)&=& \frac{(2 s_0 + \frac{1}{2}) - \lambda m^2(s_0^2-1)
                                                            (2 s_0 + 1) }
                    {4 \pi (1 + s_0) \Gamma(s_0) \Gamma(1 + s_0)}
		\Bigg[ \int dk \frac{\Gamma \left( s_0 + \frac{1}{2} 
		                                + \frac{i k}{2 m} \right)
				     \Gamma \left( s_0 + \frac{1}{2}
				                - \frac{i k}{2 m} \right)}
				     {1 + k^2 / m^2}
		         \frac{k}{m} \sin k x
			                          \\   \nonumber
           & & \hspace{5.0cm} + \tanh m x \int dk
	                        \frac{\Gamma \left( s_0 + \frac{1}{2} 
                                               + \frac{i k}{2 m} \right)
                                       \Gamma \left( s_0 + \frac{1}{2}
                                               - \frac{i k}{2 m} \right)}
                                      {1 + k^2 / m^2}
                                                           \cos k x \Bigg]
						\\   \nonumber
g_{v, 2}(x)&=& \frac{(2 s_0 + \frac{1}{2}) + \lambda m^2 (s_0^2-1)}
                    {4 \pi (1 + s_0) \Gamma(s_0) \Gamma(1 + s_0)}
	       \Bigg[ \int dk \frac{\Gamma \left( s_0 + \frac{1}{2}
	                                          + \frac{i k}{2 m}
						                  \right)
	                             \Gamma \left( s_0 + \frac{1}{2}
				                  - \frac{i k}{2 m}
						                  \right)}
                                    {(4 s_0^2 - 3) + k^2 / m^2}
			\frac{k}{m} \sin k x   \\ \nonumber
           & & \hspace{5.0cm} + \tanh m x \int dk
	                        \frac{\Gamma \left( s_0 + \frac{1}{2}
		                                   + \frac{i k}{2 m}
		                                                     \right)
		                      \Gamma \left( s_0 + \frac{1}{2}
		                                     - \frac{i k}{2 m}
		                                                     \right)}
		                      {(4 s_0^2 - 3) + k^2 / m^2}
				         \cos k x    \Bigg].
%\label{guv12}
\end{eqnarray}
By inserting Eq.(\ref{guv12}) into Eq.(\ref{gu1}) one can 
obtain double-integral 
representation of $G_{u, 1}(x)$. Then, through some appropriate change of 
variables we get the final form of $l(\Omega) - l(\Omega_{sph})$:
\begin{eqnarray}
\label{fshift2}
l(\Omega) - l(\Omega_{sph})&=& a^2 s^{-2 s_0}
                               \frac{m^3 \Gamma (1 + 2 s_0)}
			            {\Gamma^2 (s_0) \Gamma^2 (1 + s_0)}
				                        \\   \nonumber
&\times& \Bigg[
               \frac{(2 s_0 + \frac{1}{2}) - \lambda m^2 (s_0^2-1)
	                                                      (2 s_0 + 1)}
                    {2 \pi (1 + s_0)}
	       \int dy \frac{\sinh y}
	                    {\cosh^{2 s_0} y (\cosh^2 y + 
			                                4 \lambda m^2)^2}
                                                              \\ \nonumber
& & \hspace{7.0cm} \times
		K_1(y) [J_1(y) + \tanh y J_2(y)]
		                                        \\   \nonumber
& & \hspace{0.5cm} + \frac{(2 s_0 + \frac{1}{2}) + \lambda m^2
                                                             (s_0^2 - 1)}
			{2 \pi (1 + s_0)}
		\int dy \frac{\sinh y}
		             {\cosh^{2 s_0} y (\cosh^2 y + 
			                           4 \lambda m^2)^2}
						       \\  \nonumber
& & \hspace{7.0cm} \times
			K_2(y) [J_3(y) + \tanh y J_4(y)]
			                                 \\ \nonumber
& & \hspace{0.5cm} + 2^{-2 s_0} \Gamma(1 + 2 s_0)
                   \bigg[ \frac{16 s_0^2 -1}{8} 
		          \int \frac{dy}{\cosh^{4 s_0 - 2} y
			                 (\cosh^2 y + 4 \lambda m^2)}
					          \\  \nonumber
& & \hspace{4.0cm}  
			 - \frac{3(1 + 2 s_0^2)}{4}
			   \int \frac{dy}{\cosh^{4 s_0} y 
			                  (\cosh^2 y + 4 \lambda m^2)}
					                        \bigg]
								  \Bigg]
%\label{fshift2}
\end{eqnarray}
where
\begin{eqnarray}
\label{assist1}
J_1(y)&=& \int_{-\infty}^{\infty} dq
               \frac{q \mid \Gamma \left( s_0 + \frac{1 + i q}{2} \right) 
	                                                            \mid^2}
		    {1 + q^2}       \sin q y   \\  \nonumber
J_2(y)&=& \int_{-\infty}^{\infty} dq
               \frac{ \mid \Gamma \left( s_0 + \frac{1 + i q}{2} \right)
	                                                            \mid^2}
                    {1 + q^2}       \cos q y   \\  \nonumber        
J_3(y)&=& \int_{-\infty}^{\infty} dq
               \frac{q \mid \Gamma \left( s_0 + \frac{1 + i q}{2} \right)
	                                                            \mid^2}
		    {q^2 + (4 s_0^2 - 3)}           \sin q y
		                                      \\  \nonumber
J_4(y)&=& \int_{-\infty}^{\infty} dq
	       \frac{\mid \Gamma \left( s_0 + \frac{1 + i q}{2} \right)
	                                                            \mid^2}
                    {q^2 + (4 s_0^2 - 3)}   \cos q y  \\  \nonumber
K_1(y)&=& \left[ - 4 s_0 - 1 + 2(1 + 2 s_0) (s_0^2-1) \lambda m^2 \right]
                                                  \cosh^2 y
	 + \left[-4 s_0 + 3 + 2(2 s_0 -1) (s_0^2 - 1) \lambda m^2 \right]
	                                          4 \lambda m^2
						            \\  \nonumber
K_2(y)&=& \left[ -2 s_0 - \frac{1}{2} - (s_0^2 - 1) \lambda m^2 \right]
                                                  \cosh^2 y
         + \left[-2 s_0 + \frac{3}{2} - (s_0^2 - 1) \lambda m^2 \right]
	                                          4 \lambda m^2.
%\label{assist1}
\end{eqnarray}

Fig. 3 shows the result of numerical calculation of 
$l(\Omega) - l(\Omega_{sph})$ with respect to $\lambda m^2$.
From Fig. 3 we can conclude that MWS model exhibits sharp first order
winding number transition when $0 < \lambda m^2 < 0.0399$ and 
$2.148 < \lambda m^2$. 
Ref.\cite{ku97} shows first order
transition when $\lambda m^2 = 0.001$, which is in agreement with
our conclusion. 
Although Gorokhov and Blatter's method cannot determine the type of 
transition for $0.0399 < \lambda m^2 < 2.148$, it is easily
conjectured using $2 S_0 / S_{sph}$ which is shown in Fig. 4.
Here, $S_0$ is classical action of vacuum instanton and
$S_{sph} = (2\pi/\Omega_{sph}) E_{sph}$, where $E_{sph}$ is space integration
of classical Lagrangian for the static sphaleron solution, and interpreted
as barrier height in usual Minkowski  space.
Fig. 4 indicates that $2S_0 / S_{sph} < 1$ when $0 < \lambda m^2 < 0.006$
and $2.52 < \lambda m^2$, and $2S_0 / S_{sph} > 1$ when
$0.006 < \lambda m^2 < 2.52$. Hence, MWS model exhibits strong
first-order transition when $0 < \lambda m^2 < 0.006$ and 
$2.52 < \lambda m^2$, and weak first-order transition when
$0.006 < \lambda m^2 < 0.0399$ and $2.148 < \lambda m^2 < 2.52$
as shown in Fig. 5. From this fact we can conjecture that smooth second order
transition occurs at intermediate region of $\lambda m^2$.

\section{Winding number transition in MW model}
In this section we will discuss the winding number transition 
in MW model by taking $\lambda \rightarrow 0$ limit in Eq.(\ref{fshift2}). 
Using integral formula\cite{gr65} 
\begin{eqnarray}
\int_0^{\infty} dx \frac{\sin a x}{\cosh \beta x} x^{2m + 1} &=&
     (-1)^{m+1} \frac{\pi}{2 \beta} 
                \frac{\partial^{2m+1}}{\partial a^{2m+1}}
		\frac{1}{\cosh \frac{a \pi}{2 \beta}}  \\  \nonumber
\int_0^{\infty} dx \frac{\cos a x}{\cosh \beta x} x^{2m} &=&
     (-1)^m \frac{\pi}{2 \beta}
            \frac{\partial^{2m}}{\partial a^{2m}}
	        \frac{1}{\cosh \frac{a \pi}{2 \beta}},
\end{eqnarray}
we can show straightforwardly
\begin{eqnarray}
\label{assist2}
J_1(y)&=& \frac{\pi}{4} \frac{\sinh y ( 3 + 4 \cosh^2 y)}{\cosh^4 y}
                                                    \\   \nonumber
J_2(y)&=& \frac{\pi}{4} \frac{1 + 4 \cosh^2 y}{\cosh^3 y}
                                                    \\   \nonumber
J_3(y)&=& \frac{\pi}{8} \int_0^{\infty} dq
                         \frac{q (q^2 + 1) (q^2 + 9)}
			      {(q^2 + 13) \cosh \frac{\pi q}{2}}
			   \sin q y                \\   \nonumber
J_4(y)&=& \frac{\pi}{8} \int_0^{\infty} dq
                        \frac{(q^2 + 1) (q^2 + 9)}
			     {(q^2 + 13) \cosh \frac{\pi q}{2}}
			   \cos q y     \\   \nonumber
K_1(y)&=&2 K_2(y) = -9 \cosh^2 y.
%\label{assist2}
\end{eqnarray}
Hence, we can obtain $l(\Omega) - l(\Omega_{sph})$ by inserting
Eq.(\ref{assist2}) into Eq.(\ref{fshift2}). 
Since $y$-integration in this case is 
analytically solved if one uses integral formula\cite{gr65},
\begin{equation}
\int_0^{\infty} dx
     \frac{\cos a x}{\cosh^{2 n + 1} \beta x} = 
     \frac{\pi (a^2 + \beta^2) ( a^2 + 3^2 \beta^2) \cdots 
            [a^2 + (2n-1)^2 \beta^2]}
	  {2 (2n)! \beta^{2n+1} \cosh \frac{a \pi}{2 \beta}},
\end{equation}
the final form of $l(\Omega) - l(\Omega_{sph})$ in MW model is simply
\begin{equation}
l(\Omega) - l(\Omega_{sph}) = - \frac{9 a^2 m^3}{2}
                            \left[\frac{1}{70} + \frac{\pi}{4096} I \right]
\end{equation}
where
\begin{equation}
I = \int_0^{\infty} dq
    \frac{(q^2+1)^3 (q^2+9)^2}{(q^2+13) \cosh^2 \frac{\pi q}{2}}.
\end{equation}
Numerical calculation shows $l(\Omega) - l(\Omega_{sph}) = -0.161116 a^2
m^3$, which means sharp first-order transition in the full range of
parameter space. This is in agreement with result of Ref.\cite{hab96}.

\section{Conclusion}
In this paper we examine the winding number transition of the 
MW model with and without Skyrem term. For MWS model we expand the field
equation around sphaleron solution, which yields the explicit form of the
fluctuation operator. The spatial fluctuation operator is proven to be 
a kind of modified P\"{o}schl-Teller type and its number of discrete
modes is dependent on the value of $\lambda m^2$. The increment of 
$\lambda m^2$ gives rise to the increase of the number of discrete modes,
and eventually the infinite number of discrete modes arises as 
$\lambda m^2$ approaches infinity.

Following Gorokhov and Blatter we derive the sufficient condition for the
sharp first-order transition in this model, which indicates that first-order
transition occurs when $0 < \lambda m^2 < 0.0399$ and $2.148 < \lambda
m^2$. Computing $2 S_0 / S_{sph}$, we conjecture that the smooth second-order
transition occurs in the intermediate region of $\lambda m^2$.

For MW model the criterion for the sharp transition can be easily derived
by taking a $\lambda \rightarrow 0$ limit to that of MWS model.
It is shown that the MW model always exhibits a first-order winding number 
transition regardless of the value of parameter. 

MW model is frequently used as a toy model of sphaleron transition for the 
electroweak theory, and hence for the investigation of the baryon number
violating processes. Recent study\cite{fr99}, however, on the 
sphaleron transition of $SU(2)$-Higgs theory\cite{com99} has shown that 
smooth second-order sphaleron transition occurs when $6.665 < M_H/M_W < 12.03$ 
although first-order transition occurs when $M_H/M_W < 6.665$.
Hence, MW model cannot play an important role as a toy model for the 
study of sphaleron transition of electroweak theory when Higgs are very
massive than $W$ particle. This means we need another toy model which
exhibits a first-order and second-order sphaleron transition depending
on the values of parameters involved in the model. Based on our study
on the relation of number of negative modes and bifurcation point analyzed in
Ref.\cite{soo99}, we have an opinion that this may be achieved by endowing
a nontrivial topology to the space coordinate. This will be discussed 
elsewhere.

\begin{appendix}
\section{Calculation of  \lowercase{$g_{v, 1}(x)$} and \lowercase{$g_{v, 2}(x)$}}
In this appendix we will calculate $g_{v, 1}(x)$ and $g_{v, 2}(x)$ defined at
Eq.(\ref{mwsguv}).

[1] $g_{v, 1}(x)$

Using complete condition for $\hat{h}_v$, it is easy to show
\begin{equation}
\label{complete1}
\hat{h}_v^{-1} \frac{\sinh m x}{\cosh^{2 s_0+2} m x}
= \int \frac{dk}{k^2 + m^2} \mid \varphi_k > < \varphi_k \mid
  \frac{\sinh m x}{\cosh^{2 s_0+2} m x} >
%\label{complete1}
\end{equation}
where $\mid \varphi_k>$ is defined at Table I.
In Eq.(\ref{complete1}) 
$<\varphi_k \mid \frac{\sinh m x}{\cosh^{2 s_0 + 2} m x} >$
can be calculated by making use of integral formula
\begin{equation}
\int_0^{\infty} dx \frac{\cos a x}{\cosh^{\nu} \beta x} = 
\frac{2^{\nu-2}}{\beta \Gamma(\nu)}
     \Gamma \left( \frac{\nu}{2} + \frac{a i}{2 \beta} \right)
     \Gamma \left( \frac{\nu}{2} - \frac{a i}{2 \beta} \right).
\end{equation}
Inserting the result of $<\varphi_k \mid \frac{\sinh m x}
{\cosh^{2 s_0 + 2} m x} >$ into Eq.(\ref{complete1}) one can obtain
\begin{eqnarray}
g_{v, 1}(x)&=& \frac{(2 s_0 + \frac{1}{2}) - \lambda m^2(s_0^2-1)
                                                            (2 s_0 + 1) }
                    {4 \pi (1 + s_0) \Gamma(s_0) \Gamma(1 + s_0)}
                                                             \\ \nonumber
	   & & \times  \Bigg[ \int dk \frac{\Gamma \left( s_0 + \frac{1}{2} 
                                       + \frac{i k}{2 m} \right)
	                             \Gamma \left( s_0 + \frac{1}{2}
	                                       - \frac{i k}{2 m} \right)}
	                           {1 + k^2 / m^2}
	                         \frac{k}{m} \sin k x 
				                           \\ \nonumber
& & \hspace{4cm}	               + \tanh m x \int dk
		          \frac{\Gamma \left( s_0 + \frac{1}{2} 
		       + \frac{i k}{2 m} \right)
		                \Gamma \left( s_0 + \frac{1}{2}
		                  - \frac{i k}{2 m} \right)}
			      {1 + k^2 / m^2}
			                \cos k x \Bigg].
\end{eqnarray}

[2] $g_{v, 2}(x)$

Using complete condition for $\hat{h}_v$ again one can show directly
\begin{eqnarray}
\label{complete2}
& &    \left[ l(2 \Omega_{sph}) - \hat{h}_v  \right]^{-1}
\frac{\sinh m x}{\cosh^{2 s_0 + 2} m x}
                                                  \\  \nonumber
&=&
-\frac{s^{2 s_0}}{4 \pi (1 + s_0) m^3 \Gamma (1 + 2 s_0)}
\int dk \frac{\Gamma \left(s_0 + \frac{1}{2} + \frac{i k}{2 m} 
                                                              \right)
                       \Gamma \left( s_0 + \frac{1}{2} - \frac{i k}{2 m}
		                                               \right)}
                      {(4 s_0^2 - 3) + k^2 / m^2}
		      \left( \frac{k}{m} \sin k x
		             + \tanh m x \cos k x  \right).
%\label{complete2}
\end{eqnarray}
Inserting Eq.(\ref{complete2}) into Eq.(\ref{mwsguv}) we obtain
\begin{eqnarray}
g_{v, 2}(x)&=& \frac{(2 s_0 + \frac{1}{2}) + \lambda m^2 (s_0^2-1)}
                    {4 \pi (1 + s_0) \Gamma(s_0) \Gamma(1 + s_0)}
	            \Bigg[ \int dk \frac{\Gamma \left( s_0 + \frac{1}{2}
	                + \frac{i k}{2 m}
	                     \right)
	               \Gamma \left( s_0 + \frac{1}{2}
	                 - \frac{i k}{2 m}
	                    \right)}
	            {(4 s_0^2 - 3) + k^2 / m^2}
	              \frac{k}{m} \sin k x   \\
	       & & \hspace{5.0cm} + \tanh m x \int dk
	          \frac{\Gamma \left( s_0 + \frac{1}{2}
	              + \frac{i k}{2 m}
	                  \right)
	     \Gamma \left( s_0 + \frac{1}{2}
	         - \frac{i k}{2 m}
	                     \right)}
	          {(4 s_0^2 - 3) + k^2 / m^2}
	         \cos k x   \nonumber \Bigg].
\end{eqnarray}
 In MW model($\lambda \rightarrow 0$, $s_0 \rightarrow 2$) $g_{v, 1}(x)$
 and $g_{v, 2}(x)$ are simply expressed as follows:
 \begin{eqnarray}
 g_{v, 1}(x)&=& \frac{3 m}{16}
                \frac{\sinh m x}{\cosh^4 m x}
		(1 + 2 \cosh^2 m x)
		                          \\   \nonumber
g_{v, 2}(x)&=& \frac{3}{256 m^2}
              \left[\frac{1}{m} \int dk 
	                   \frac{k (k^2 + m^2) (k^2 + 9 m^2)}
			        {(k^2 + 13 m^2) \cosh
				   \frac{k \pi}{2 m} }
		            \sin k x 
		     + \tanh m x 
		       \int dk \frac{(k^2 + m^2)(k^2 + 9 m^2)}
		                    {(k^2 + 13 m^2) \cosh 
				                   \frac{k \pi}{2 m} }
			    \cos k x
			                               \right].
\end{eqnarray}

\end{appendix}

\vspace{5mm}
\begin{table}
%\centerline{\bf Table I}
%\centerline{eigenvalues and eigenfunctions of $\hat{h}_v$}
%\vspace{2mm}
\begin{tabular}{|c|c|c|}
%\hline
$     $ & $\epsilon_n^v$ \hspace{2.5cm} & $\varphi_n(x)$ \hspace{2cm} \\
                                       \hline
$n = 0$ & $0 $\hspace{2.5cm} & $\sqrt{\frac{m}{2}} \frac{1}{\cosh m x}$ \hspace{2cm} \\ \hline
$n = k$ & $m^2 + k^2$ \hspace{2.5cm} & $\frac{1}{\sqrt{2 \pi}} \frac{e^{i k x}}
                                                      {1 + \frac{i k}{m}}
                           \left( \frac{ik}{m} - \tanh m x \right)$\hspace{2cm} \\ 
%\hline
\end{tabular}

\vspace{0.5cm}
 
\caption{eigenvalues and eigenfunctions of $\hat{h}_v$}

\vspace{2cm}

\begin{tabular}{|c|c|c|}
$ $&$\epsilon_n^u$&$\psi_n(x)$  \\  \hline
$n = 0$&$m^2 (1 - s_0^2)$&$\sqrt{\frac{m \Gamma(1 + 2 s_0)}
                                           {\Gamma(s_0) \Gamma(1 + s_0)}}
                           \left( \frac{1}{2 \cosh m x} \right)^{s_0}$ \\
                                                                 \hline
$n = 1$& $0$ &$\sqrt{\frac{3 m}{2}} \frac{\sinh m x}
                                              {\cosh^2 m x}$ \\ \hline
$n = 2$&$m^2[1 - (s_2 - 2)^2]$&$\sqrt{\frac{m \Gamma(2 s_2 - 1)}
                                                {2 \Gamma(s_2-1)\Gamma(s_2-2)}}
                                \left( \frac{1}{2 \cosh m x}
                                                             \right)^{s_2-2}
                                \frac{(2s_2-1) \tanh^2 m x -1}
                                     {2(s_2-1)}$   \\   \hline
$n = k$&$m^2 + k^2$&$\frac{1}{\sqrt{2 \pi}}
                       \frac{\Gamma \left( -s_k - \frac{i k}{m} \right)
                            \Gamma \left( s_k + 1 - \frac{i k}{m} \right)}
                            {\Gamma \left( -\frac{i k}{m} \right)
                             \Gamma \left( 1 - \frac{i k}{m} \right)}
                       e^{i k x} \hspace{2mm}_2F_1
                                     \left( s_k+1, -s_k; 1 - \frac{i k}{m};
                                          \frac{1}{2} (1 - \tanh m x)
                                                                  \right)$ \\
\end{tabular}

\vspace{0.5cm}

\caption{eigenvalues and eigenfunctions of $\hat{h}_u$ when 
         $0 < \lambda m^2 < \frac{3}{2}$}
\end{table}

\begin{figure}

\caption{ Typical Action-vs-Period diagrams of first-order transition for (a) case I
         and (b) case II. Bold lines determine the winding number transitions.}
\end{figure}

\begin{figure}

\caption{Discrete spectrum of $\hat{h}_u$. The number of discrete modes
is dependent on the value of $\lambda m^2$. In fact, there are $n$
discrete modes when $\frac{(n-1)^2 + (n-1) - 6}{4} < \lambda m^2
\leq \frac{n^2 + n - 6}{4}$}.
\end{figure}

\begin{figure}

\caption{$\lambda m^2$-dependence of $l(\Omega) - l(\Omega_{sph})$.
This figure shows that MWS model exhibits sharp first-order transition
when $0 < \lambda m^2 < 0.0399$ and $2.148 < \lambda m^2$.}
\end{figure}

\begin{figure}

\caption{$\lambda m^2$-dependence of $2S_0 / S_{sph}$.
Combining this figure and Fig.3 we can conclude that MWS model exhibits
weak first-order transition when $0.006 < \lambda m^2 < 0.0399$
and $2.148 < \lambda m^2 < 2.52$. From this fact one can conjecture
MWS model exhibits smooth second-order transition when $0.0399 < \lambda 
m^2 < 2.148$.}
\end{figure}

\begin{figure}
\caption{(a) Action-vs-Period diagram of MWS model when $0 < \lambda m^2
< 0.006$ and $2.52 < \lambda m^2$. In this regime MWS model exhibits
strong first-order sphaleron transition.
(b) Action-vs-Period diagram of MWS model when $0.006 < \lambda m^2
< 0.0399$ and $2.148 < \lambda m^2 < 2.52$. In this regime MWS model
exhibits weak first-order sphaleron transition.
(c) Action-vs-Period diagram of MWS model when $0.0399 < \lambda m^2 <
2.148$, which is drawn by conjecture from the fact that MWS model 
exhibits weak first-order transition when $0.006 < \lambda m^2 <
0.0399$ and $2.148 < \lambda m^2 < 2.52$.}
\end{figure}
\epsfysize=25cm \epsfbox{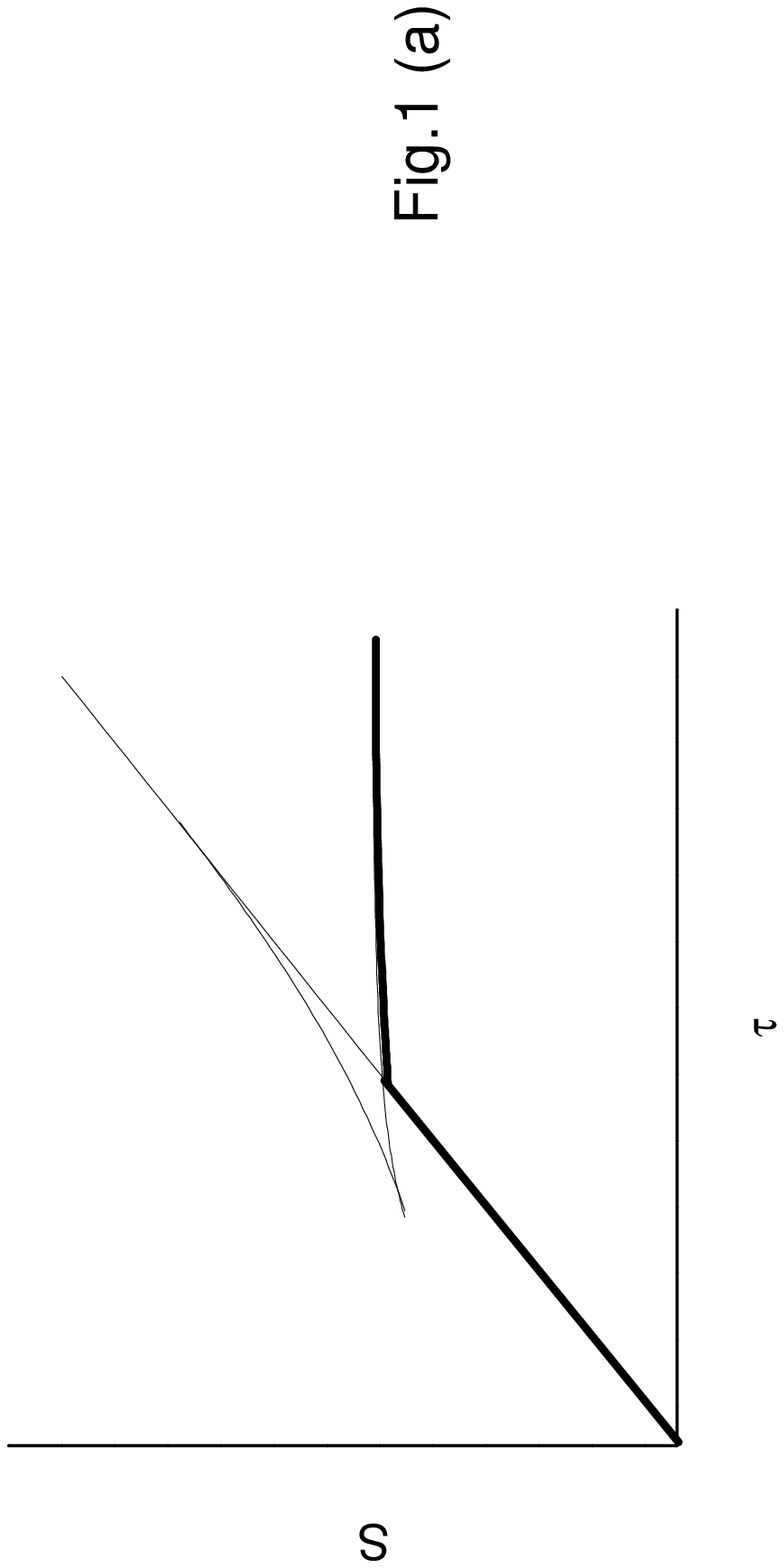}
\epsfysize=25cm \epsfbox{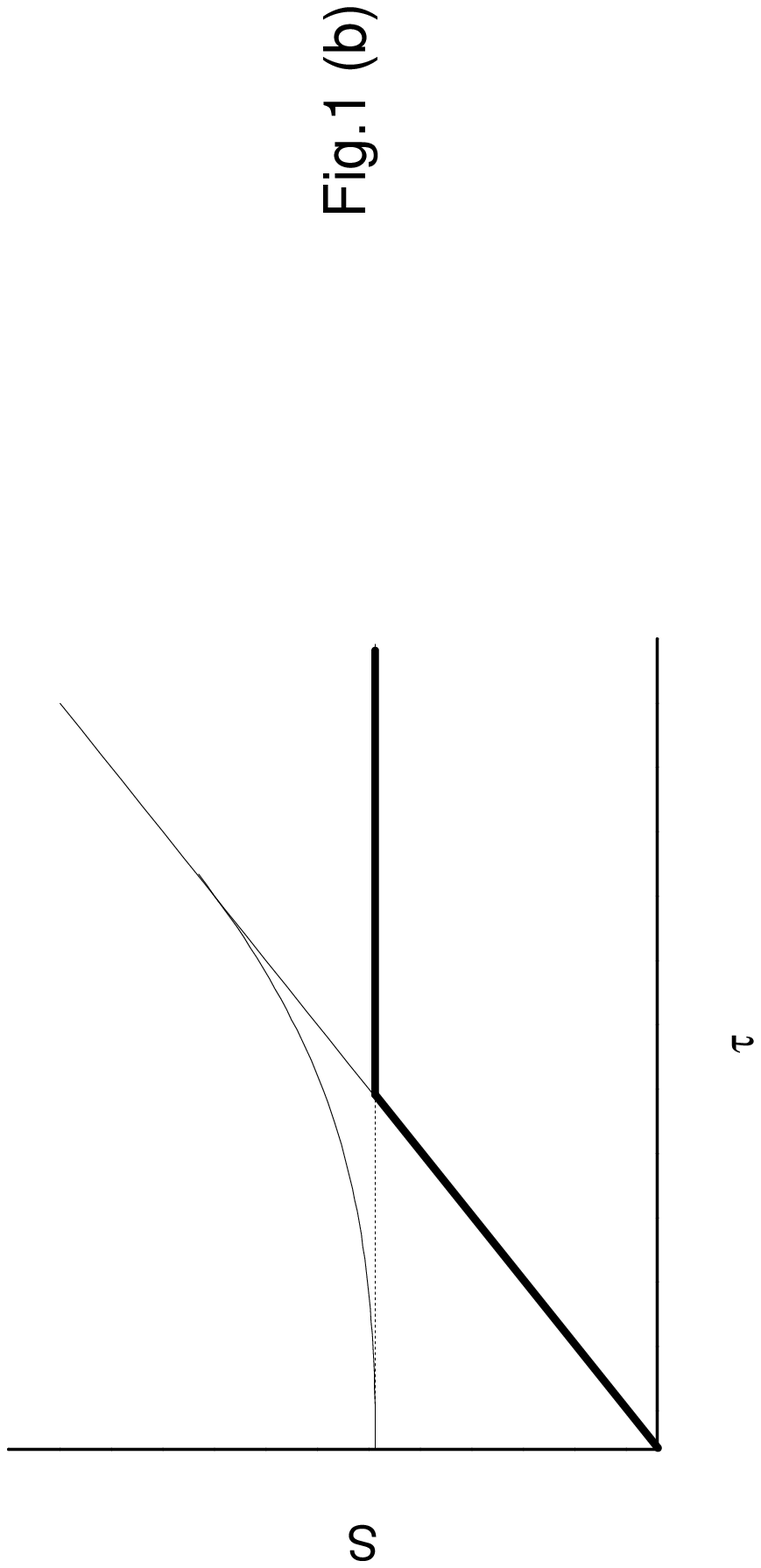}
\epsfysize=25cm \epsfbox{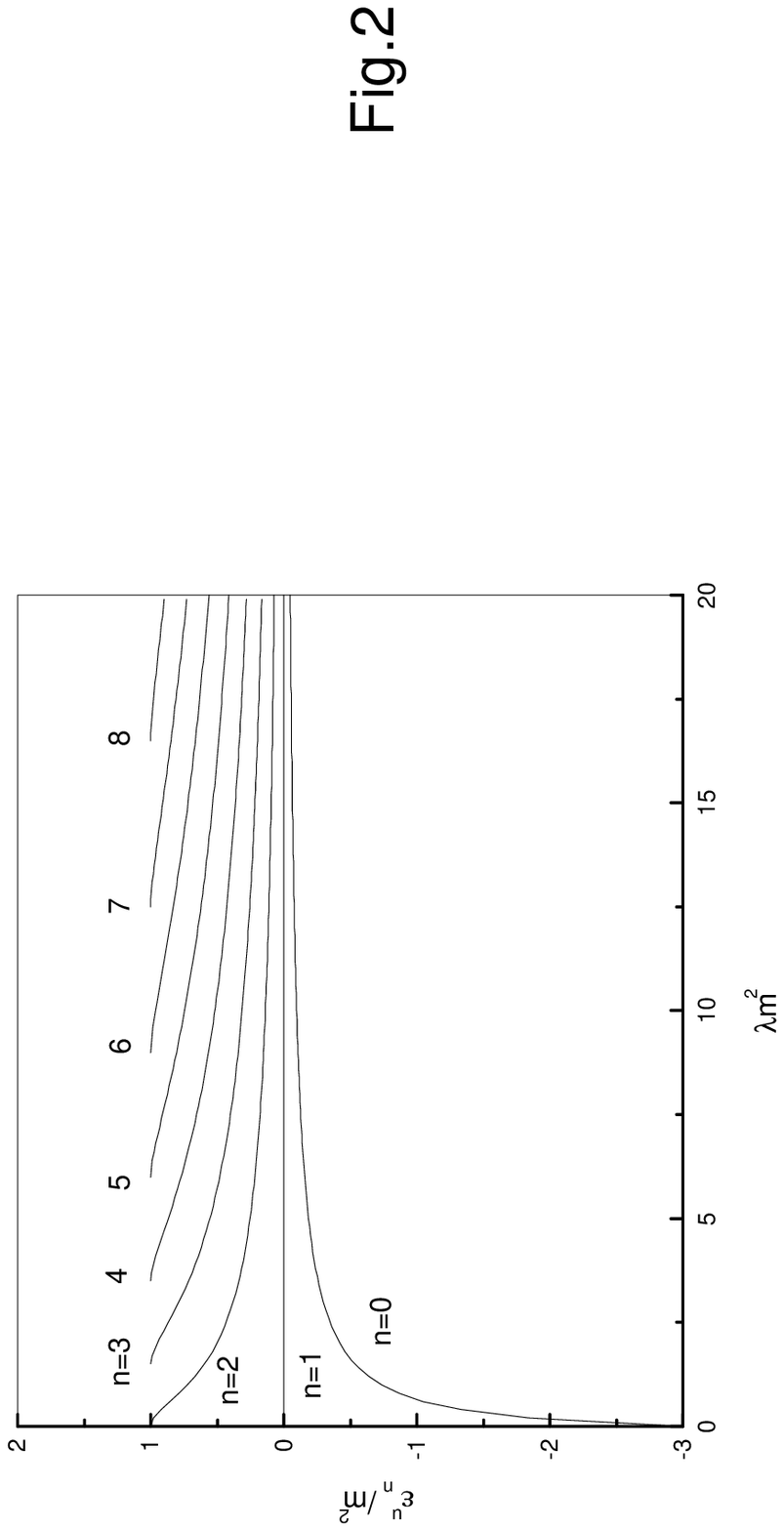}
\epsfysize=25cm \epsfbox{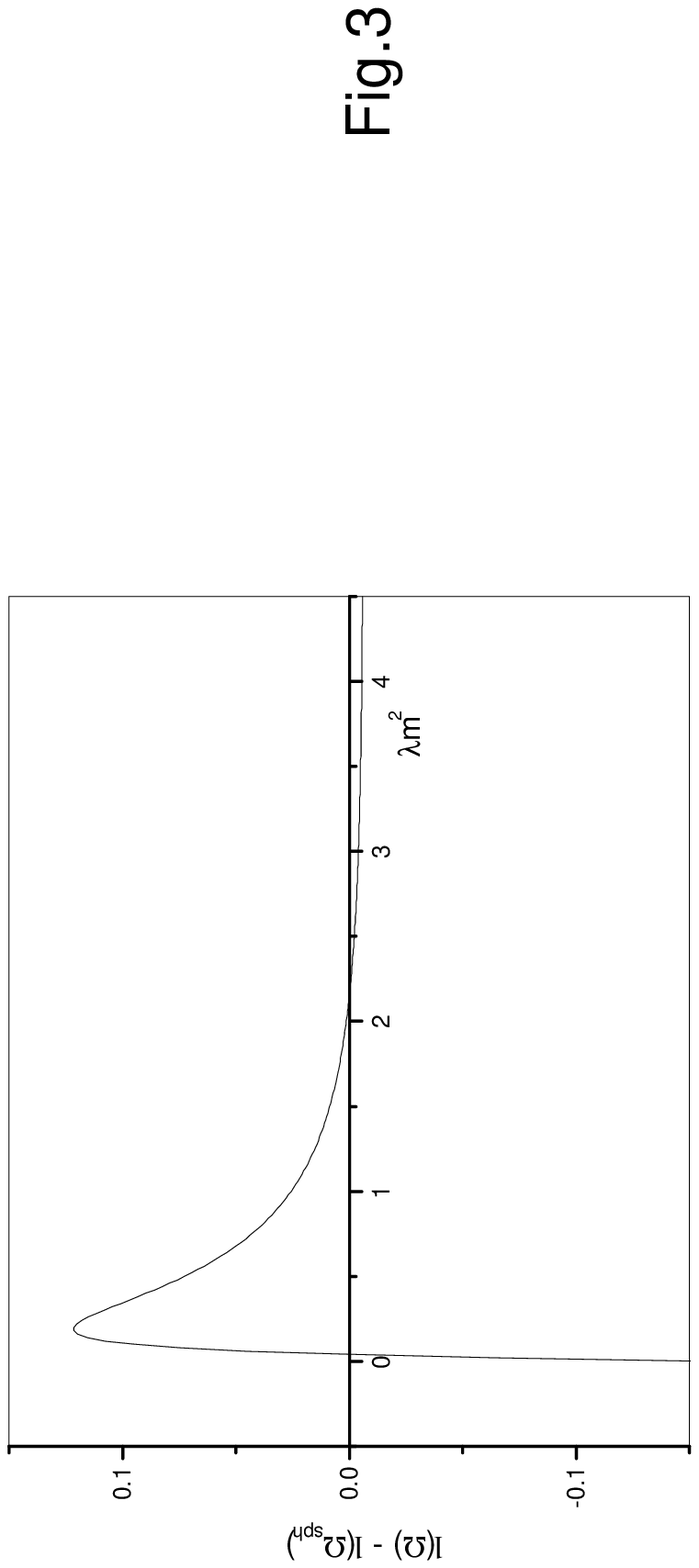}
\epsfysize=25cm \epsfbox{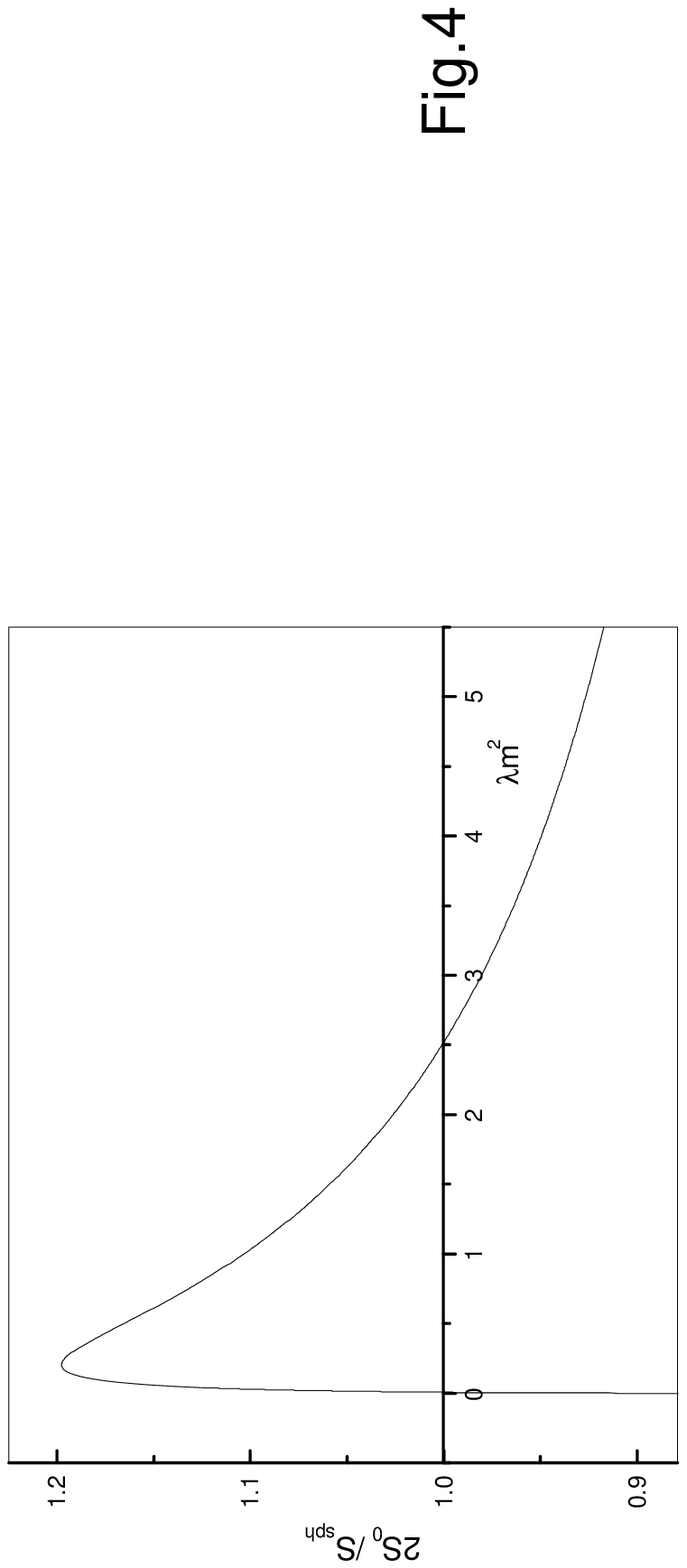}
\epsfysize=25cm \epsfbox{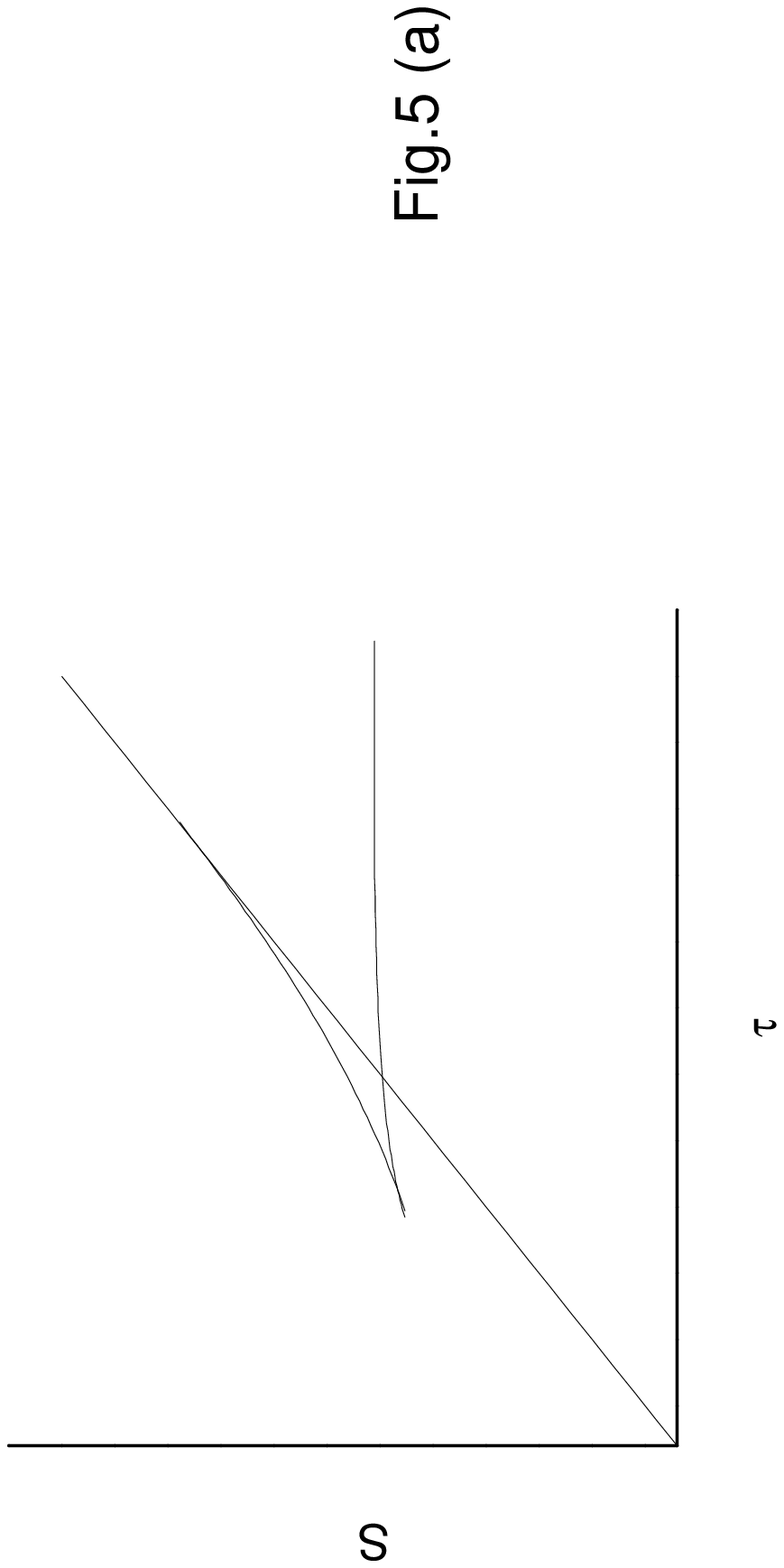}
\epsfysize=25cm \epsfbox{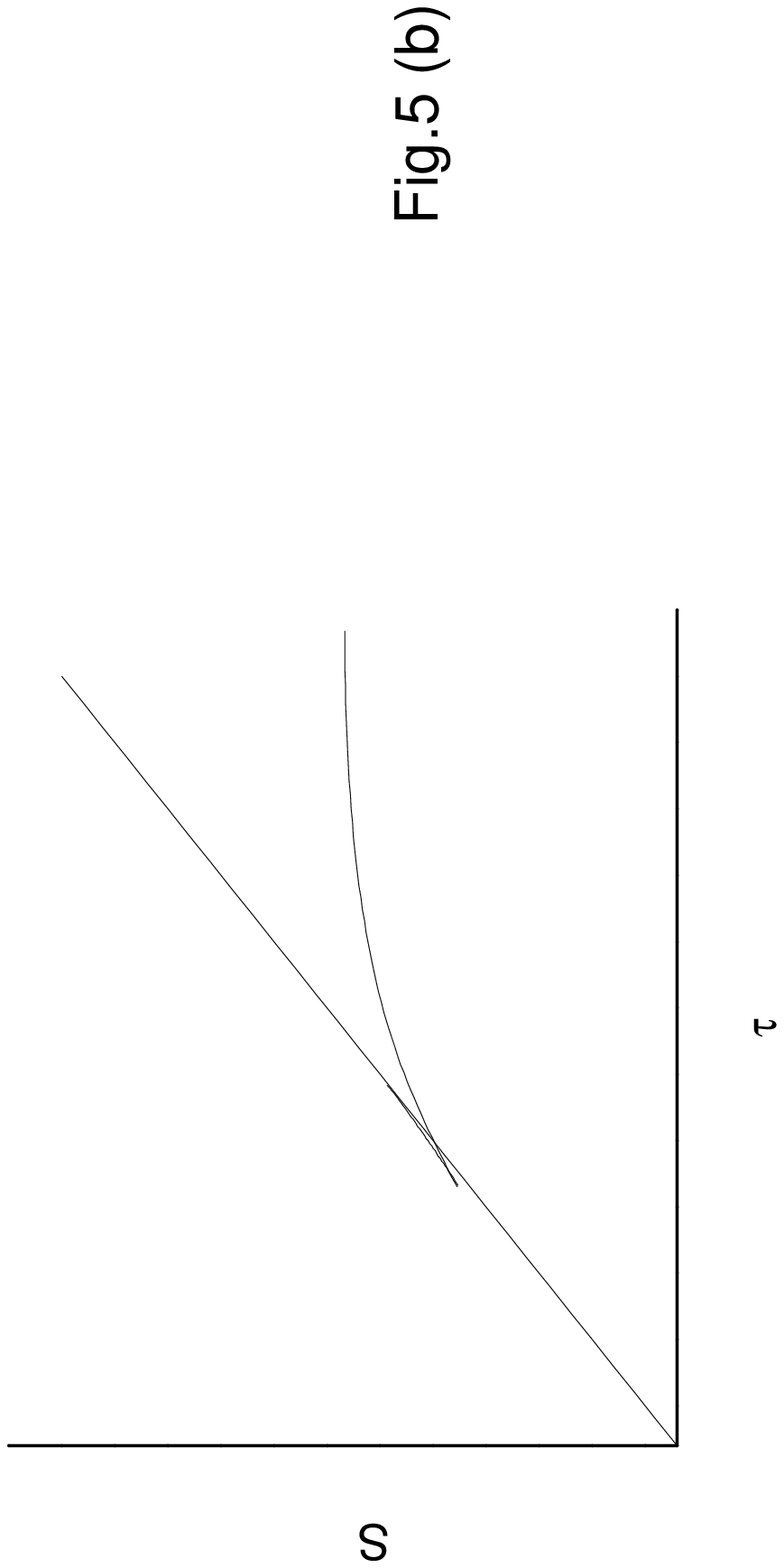}
\epsfysize=25cm \epsfbox{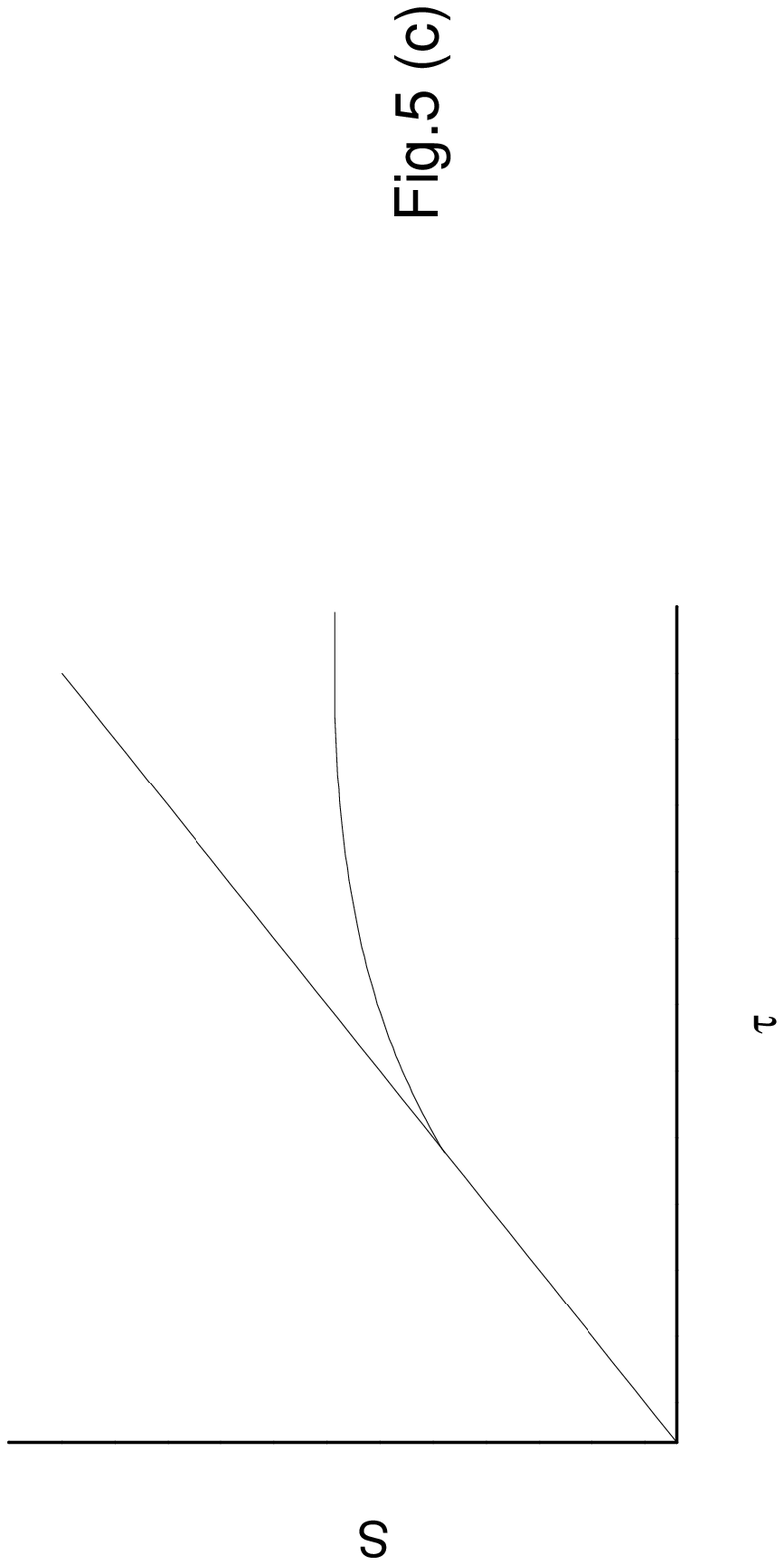}


\begin{thebibliography}{99}
\bibitem{com99} $SU(2)$-Higgs theory is a bosonic sector of electroweak 
theory in the limit of vanishing weak mixing angle. See Ref.\cite{ya89}.
\end{thebibliography}
\end{document}